\documentclass{article}
\usepackage[margin=1in]{geometry}
\usepackage{amsmath}
\usepackage{epsfig}
\numberwithin{equation}{section} 
\newcommand{\beq}{\begin{equation}}
\newcommand{\eeq}{\end{equation}}
\newcommand{\beqa}{\begin{eqnarray}}
\newcommand{\eeqa}{\end{eqnarray}}
\newcommand{\bdm}{\begin{displaymath}}
\newcommand{\edm}{\end{displaymath}}
\newcommand{\lslash}[1]{#1\llap/}

\newcommand{\Eq}[1]{Eq.\ (\ref{#1})}
\newcommand{\Eqs}[2]{Eqs.\ (\ref{#1}) and (\ref{#2})}
\newcommand{\Eqss}[3]{Eqs.\ (\ref{#1}), (\ref{#2}) and (\ref{#3})}

\newcommand{\Ref}[1]{Ref.\ \cite{#1}}
\newcommand{\Fig}[1]{Fig.\ \ref{#1}}
\newcommand{\Table}[1]{Table\ \ref{#1}}
\newcommand{\Tr}{\mbox{Tr}\,}
\newcommand{\Section}[1]{Section\ \ref{#1}}
\newcommand{\Appendix}[1]{Appendix\ \ref{#1}}
\newcommand{\fEp}{f}
\newcommand{\fEpp}{f^\prime}
\newcommand{\fbEp}{\bar f}
\newcommand{\fbEpp}{\bar f^\prime}
\newcommand{\fnu}{f^\prime_\nu}
\newcommand{\fnub}{\bar f^\prime_\nu}

\title{
  Neutrino decoherence in a fermion and scalar background
}
  
\author{Jos\'e F. Nieves\footnote{nieves@ltp.uprrp.edu}\\
  Laboratory of Theoretical Physics, Department of Physics\\
  University of Puerto Rico, R\'{\i}o Piedras, Puerto Rico 00936
  \and\\[12pt]
  Sarira Sahu\footnote{sarira@nucleares.unam.mx}\\
  Instituto de Ciencias Nucleares\\
  Universidad Nacional Aut\'onoma de Mexico\\
  Circuito Exterior, C. U.\\
  A. Postal 70-543, 04510 Mexico DF, Mexico\\
}
\date{}

\begin{document}
\maketitle

\begin{abstract}
We consider the decoherence effects in the propagation of neutrinos
in a background composed of a scalar particle and a fermion
due to the non-forward neutrino scattering processes.
Using a simple model for the coupling of the form $\bar f_R\nu_L\phi$ 
we calculate the contribution to the imaginary part of the neutrino
self-energy arising from the non-forward neutrino scattering processes
in such backgrounds, from which the damping terms are determined.
In the case we are considering, in which the initial neutrino state is
depleted but does not actually disappear (the initial neutrino
transitions into a neutrino of a different flavor but does not decay
into a $f\phi$ pair, for example),
we associate the damping terms with decoherence effects.
For this purpose we give a precise
prescription to identify the decoherence terms, as used in the context
of the master or Linblad equation, in terms of the damping terms
we have obtained from the calculation of the imaginary part of the neutrino
self-energy from the non-forward neutrino scattering processes.
The results can be directly useful in the context of Dark Matter-neutrino
interaction models in which the scalar and/or fermion constitute
the dark-matter, and can also serve to guide the generalizations
to other models and/or situations in which the decoherence
effects in the propagation of neutrinos originate from the non-forward
scattering processes may be important. As a guide to
estimating such decoherence effects, the contributions to the
absorptive part of the self-energy and the corresponding damping terms
are computed explicitly in the context of the model we consider,
for several limiting cases of the momentum distribution functions
of the background particles.
\end{abstract}

\section{Introduction and Summary}
\label{sec:introduction}

Several extensions of the standard electroweak theory
involve the coupling of neutrinos to scalar particles ($\phi$)
and fermions ($f$) of the generic form $\bar f_R\nu_L\phi$.
Such couplings have been considered recently in the context of
Dark Matter-neutrino interactions\cite{Mangano:2006mp,Binder:2016pnr,
Primulando:2017kxf,Campo:2017nwh,Brune:2018sab,Franarin:2018gfk,Dev:2019anc,
Pandey:2018wvh,Karmakar:2018fno}.
Those interactions can produce nonstandard contributions to the neutrino index
of refraction and effective potential when the neutrino propagates
in a background of those particles.
In \Ref{Nieves:2018vxl} we considered the real part of the self-energy
of a neutrino that propagates in a medium consisting of fermions and scalars,
with a coupling of that form.
From the self-energy, the neutrino and antineutrino effective potential and
dispersion relations were then determined.

In the presence of these interactions there can also be damping terms
in the neutrino effective potential and index of refraction,
as a consequence of processes such as
$\nu + \phi \leftrightarrow f$ and $\nu + \bar f\leftrightarrow \bar\phi$,
that may become possible depending on the kinematics conditions.
In \Ref{nsnuphidamp}, we extended our previous work
to calculate the imaginary part (or more precisely the absorptive part)
of the neutrino self-energy, in a fermion and scalar background
due to the $\bar f_R\nu_L\phi$ interaction. From the imaginary part of the 
self-energy the damping terms in the effective potential and dispersion
relation were obtained.

Here we note that, in addition to the effects we have mentioned,
the presence of those couplings in general can induce decoherence effects
in the propagation of neutrinos
due to the neutrino non-forward scattering process
\cite{Coloma:2018idr,Carpio:2017nui,Oliveira:2014jsa,
Fogli:2007tx,Capolupo:2018hrp}. To be more precise,
here we consider various neutrino flavors ($\nu_{La}$) interacting with
a scalar and fermion with a coupling of the form
\beq
\label{Lfnuphia}
L_{int} = \sum_a \lambda_a\bar f_R \nu_{La} \phi + h.c.
\eeq
In this case, the scattering processes of the form
$\nu_a + x \rightarrow \nu_b + x$, where $x = f,\phi$,
can induce decoherence effects in the propagation of neutrinos,
independently of the possible damping effects already mentioned.
From the calculational point of view, the first step in our strategy is to
determine the contribution of such processes to the absorptive part of the
self-energy, from which we obtain the corresponding contribution to the
damping matrix $\Gamma$ by the usual method. However,
in the present case, in which the initial neutrino state is
depleted but does not actually disappear (the initial neutrino transitions
into a neutrino of a different flavor but does not decay into a $f\phi$ pair,
for example), the effects of the non-forward scattering
processes are more properly interpreted in terms of decoherence phenomena
rather than damping. The second step in our strategy is to give 
a precise prescription to identify the decoherence terms, as used in the context
of the master or Linblad equation, in terms of the damping matrix $\Gamma$
that we obtain from the calculation of the imaginary part of the neutrino
self-energy due to the non-forward neutrino scattering processes.
In writing \Eq{Lfnuphia} we assume
the presence of only one scalar and one fermion field.
Despite this simplification our work illustrates some features that can
serve as a guide when considering more general cases
or situations not envisioned here. They can be applied, for example,
in the context of models in which sterile ($\nu_{Ls}$)
neutrinos have \emph{secret} gauge interactions of the form
$\bar\nu_{Ls}\gamma^\mu \nu_{Ls} A^\prime_\mu$ \cite{Chu:2018gxk},
when a sterile neutrino propagates in a background of sterile
neutrinos and $A^\prime$ bosons.  They can also be applied
in models in which sterile neutrinos interact with the active neutrinos
via coupling of the form
$\lambda_a {\bar \nu}^c_{Rs} {\nu}_{La}\phi$\cite{Primulando:2017kxf,
Franarin:2018gfk}.
The formulas we obtain for the damping and decoherence terms
can be applied in the context of such models with minor modifications.
As usual, the formulas involve integrals over the momentum distribution
functions of the background particles.
As a guide to estimating such decoherence effects, the contributions to the
absorptive part of the self-energy and the corresponding damping terms
are computed explicitly in the context of the model we consider,
for several limiting cases of the momentum distribution functions
of the background particles.

In \Section{sec:dampingmatrix} we review the method we used previously
to determine the dispersion relation and damping term for a single
neutrino generation propagating in an $f\phi$ background,
and then extend it here to the case of several neutrino generations,
in particular to determine the damping matrix from the calculation
of the self-energy. In \Section{sec:sigma12} we carry out the
calculation of the absorptive part of the self-energy that arises
from the non-forward neutrino scattering processes. For definiteness
we consider the special situation in which there are no $\phi$ scalars
in the background (\emph{the heavy $\phi$ limit}), so the background
consists of the $f$ fermions and the antiparticles only. The final
result in that section is the formula for the damping matrix, expressed
in terms of integrals over the background fermion distribution functions
and the coupling constants $g_a$ defined in \Eq{Lfnuphia}.
In \Section{sec:decoherence} we formulate the interpretation
of the damping matrix so determined as a decoherence effect and its
relation to the Linblad equation and the stochastic evolution
of the state vector\cite{Daley:2014fha,Weinberg:2011jg,
pearle,Plenio:1997ep,Lieu:2019cev}.
The result is a well-defined formula
for the ``jump'' operators in that context. In \Section{sec:examples}
the integrals involved are evaluated explicitly for different
conditions of the fermion background.
Our conclusions and outlook are given in \Section{sec:conclusions},
and some details of the derivations are provided in the Appendix.

\section{Self-energy and the damping matrix}
\label{sec:dampingmatrix}

\subsection{Dispersion relation for a single neutrino generation}

\begin{figure}
\begin{center}
\epsfig{file=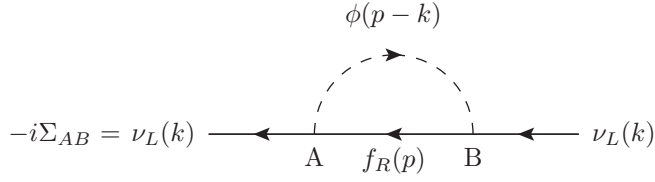,bbllx=166,bblly=366,bburx=404,bbury=431}
\end{center}
\caption{
  One-loop diagram for the neutrino thermal self-energy matrix elements
  in an $f\phi$ background.
  \label{fig:oneloop}
}
\end{figure}

In order to set down our notation and conventions it is useful
to first review briefly the case of only one neutrino generation coupled in
\Eq{Lfnuphia}, considered in Refs.\ \cite{Nieves:2018vxl,nsnuphidamp}.
We denote by $u^\mu$ the velocity four-vector of the
background medium and by $k^\mu$ the momentum of the
propagating neutrino. In the background medium's own rest frame,
\beq
\label{restframe}
u^\mu = (1,\vec 0)\,,
\eeq
and in this frame we also write
\beq
\label{krestframe}
k^\mu = (\omega,\vec\kappa)\,.
\eeq
In this work we consider only one background medium, which
can be taken to be at rest, and therefore we adopt \Eqs{restframe}{krestframe}
throughout. The dispersion relation $\omega(\vec\kappa)$ and the spinor of
the propagating mode are determined by solving the equation
\beq
\label{eveq}
\left(\lslash{k} - \Sigma_{eff}\right)\psi_L(k) = 0\,,
\eeq
where $\Sigma_{eff}$ is the neutrino thermal self-energy. $\Sigma_{eff}$
can be decomposed in the form
\beq
\Sigma_{eff} = \Sigma_r + i\Sigma_i\,,
\eeq
where $\Sigma_r$ is the dispersive part and $\Sigma_i$ the absorptive part,
\beqa
\label{Sigmaridef}
\Sigma_r & = & \frac{1}{2}(\Sigma_{eff} + \overline\Sigma_{eff})\,,\nonumber\\
\Sigma_i & = & \frac{1}{2i}(\Sigma_{eff} - \overline\Sigma_{eff})\,,
\eeqa
with
\beq
\overline\Sigma_{eff} = \gamma^0\Sigma_{eff}\gamma^0\,.
\eeq
In the context of thermal field theory
\beq
\Sigma_r = \Sigma_{11r} \equiv
\frac{1}{2}(\Sigma_{11} + \overline\Sigma_{11})\,,
\eeq
where $\Sigma_{11}$ is the $11$ element of the thermal self-energy matrix.
On the other hand, $\Sigma_i$ is conveniently obtained from the formula
\beq
\label{Sigmaidef1}
\Sigma_i = \frac{\Sigma_{12}}{2i n_F(x_\nu)}\,,
\eeq
where $\Sigma_{12}(k)$ is the $12$ element of the neutrino
thermal self-energy matrix, while
\beq
\label{nzfermion}
n_F(z) = \frac{1}{e^z + 1}\,,
\eeq
is the fermion distribution function, written in terms of a dummy variable $z$,
and the variable $x_\nu$ is given by
\beq
x_\nu = \beta k\cdot u - \alpha_\nu\,.
\eeq
To the lowest order, $\Sigma_{11}$ and $\Sigma_{12}$ are determined
by evaluating the diagram shown in \Fig{fig:oneloop}.
The chirality of the neutrino interactions imply that\cite{footnote1}
%
\beq
\label{SigmaV}
\Sigma = V^\mu(\omega,\vec\kappa)\gamma_\mu L\,,
\eeq
and correspondingly
\beq
\label{SigmariVri}
\Sigma_{r,i} = V^\mu_{r,i}(\omega,\vec\kappa)\gamma_\mu L\,,
\eeq
with
\beq
\label{Vri}
V^\mu = V^\mu_r + iV^\mu_i\,.
\eeq
We have indicated explicitly that, in general, both $V^\mu_{r,i}$ are
functions of $\omega$ and $\vec\kappa$.
Ordinarily we will omit those arguments but we will restore them when needed.

The results obtained in \Ref{nsnuphidamp} are summarized as follows.
Writing the neutrino and antineutrino dispersion relations in the form
\beq
\label{disprelform}
\omega^{(\nu,\bar\nu)} = \omega^{(\nu,\bar\nu)}_r -
\frac{i\gamma^{(\nu,\bar\nu)}}{2}\,,
\eeq
$\omega^{(\nu,\bar\nu)}_r$ is given by
\beq
\label{nudisprelreal}
\omega^{(\nu,\bar\nu)}_r = \kappa + V^{(\nu,\bar\nu)}_{eff}
\eeq
where $V^{(\nu,\bar\nu)}_{eff}$ are the effective potentials
\beqa
\label{Veff}
V^{(\nu)}_{eff} & = & n\cdot V_r(\kappa,\vec\kappa) =
 V^0_r(\kappa,\vec\kappa) - \hat\kappa\cdot\vec V_r(\kappa,\vec\kappa)
\,,\nonumber\\
V^{(\bar\nu)}_{eff} & = & -n\cdot V_r(-\kappa,-\vec\kappa) =
 -V^0_r(-\kappa,-\vec\kappa) + \hat\kappa\cdot\vec V_r(-\kappa,-\vec\kappa)\,,
\eeqa
with
\beq
\label{nmu}
n^\mu = (1,\hat\kappa)\,.
\eeq
On the other hand, for the imaginary part,
\beqa
\label{nudisprelimg}
-\frac{\gamma^{(\nu)}(\vec\kappa)}{2} & = & 
\frac{n\cdot V_i(\kappa,\vec\kappa)}
{1 - n\cdot\left.
\frac{\partial V_r(\omega,\vec\kappa)}{\partial\omega}
\right|_{\omega = \kappa}}\,,\nonumber\\
-\frac{\gamma^{(\bar\nu)}(\vec\kappa)}{2} & = & 
\frac{n\cdot V_i(-\kappa,-\vec\kappa)}
{1 - n\cdot\left.
\frac{\partial V_r(\omega,-\vec\kappa)}{\partial\omega}
\right|_{\omega = -\kappa}}\,,
\eeqa
where $n^\mu$ is defined in \Eq{nmu}.
If the correction due to the
$n\cdot\partial V_r(\omega,\vec\kappa)/\partial\omega$ in the denominator
can be neglected, the formulas in \Eq{nudisprelimg} reduce to
\beqa
\label{nudisprelimg-simple}
-\frac{\gamma^{(\nu)}(\vec\kappa)}{2} & = & 
n\cdot V_i(\kappa,\vec\kappa)\,,\nonumber\\
-\frac{\gamma^{(\bar\nu)}(\vec\kappa)}{2} & = & 
n\cdot V_i(-\kappa,-\vec\kappa)\,.
\eeqa
In any case, \Eqs{nudisprelreal}{nudisprelimg}, allow us
to obtain the neutrino and antineutrino dispersion relation and
damping from the self-energy.

\subsection{Several generations - equation for the flavor spinors}

Our aim here is to extend the above considerations to
the case of several neutrino generations.
In this case $V^\mu_{r,i}$, as well as $\Sigma_{r,i}$,
are matrices in flavor space.
As already stated, in this work we assume that the distribution functions
of the background particles are isotropic.
In this case $V^\mu$ is a function only
of $k^\mu$ and $u^\mu$ and no other vectors. One traditional way
to take this into account is to parameterize $V^\mu$ in the form
\beq
V^\mu = ak^\mu + bu^\mu\,.
\eeq
For our present purposes we find more convenient to proceed as follows.
The isotropy assumption is equivalent to assume that
\beq
\label{Vrestframe}
V^\mu = (V^{(u)},V^{(t)}\hat\kappa)\,,
\eeq
in that frame. For completeness we note that this can be written in
a general way by introducing
\beq
t^\mu = \frac{1}{\kappa}(k^\mu - \omega u^\mu)\,,
\eeq
where
\beqa
\omega & = & k\cdot u\,,\nonumber\\
\kappa & = & \sqrt{\omega^2 - k^2}\,,
\eeqa
and
\beq
V^\mu = V^{(u)} u^\mu + V^{(t)} t^\mu\,,
\eeq
In what follows we adopt the conventions defined by 
\Eqss{restframe}{krestframe}{Vrestframe} throughout.

Our job here is to find out what is the Hamiltonian for the evolution
equation of the flavor amplitudes. We do it in the following steps:
\begin{enumerate}
\item We write the $\psi$ field in \Eq{eveq} schematically
  as the product $u\xi$, where $u$ is a Dirac spinor and $\xi$ is a flavor
  spinor.

\item From \Eq{eveq} we obtain an equation for $\xi$ and the dispersion
  relation, leaving the Dirac matrix structure behind.

\item We then express the equation for $\xi$ and the dispersion relation
  in the form
  \beq
  H\xi = \omega\xi
  \eeq
  which will identify the Hamiltonian.
\end{enumerate}
The details follow.
  
As in the case of one generation, \Eq{eveq} has positive and negative
frequency solutions. To distinguish them, we use the superscripts
$\lambda = \pm$ on the relevant quantities. We introduce the
  \emph{positive} ($u^{(+)}_L(\vec\kappa)$) and
  \emph{negative} ($u^{(-)}_L(\vec\kappa)$)
\emph{frequency} left-handed chiral Dirac spinor satisfying
\beq
\label{uL}
\lslash{n}^{(\lambda)} u^{(\lambda)}_L = 0\,,
\eeq
with
\beq
\label{nhat}
n^{(\lambda)\mu} = (1,\lambda\hat\kappa)\,.
\eeq
To solve \Eq{eveq} we write the ansatz
\beq
\label{ansatz}
\psi^{(\lambda)}_L = u^{(\lambda)}_L(\vec\kappa)\xi^{(\lambda)}(\vec\kappa)\,,
\eeq
where $\xi^{(\lambda)}$ is a flavor spinor, representing the amplitude
in flavor space. \Eq{uL} implies that
\beqa
\label{uLrelations}
\lslash{k} u^{(\lambda)}_L & = & (\omega - \lambda\kappa)\lslash{u}
u^{(\lambda)}_L\,,\nonumber\\
\lslash{V}L u^{(\lambda)}_L & = & (V^{(u)} - \lambda V^{(t)})\lslash{u}
u^{(\lambda)}_L\,.
\eeqa
Substituting \Eq{ansatz} in \Eq{eveq} and using \Eq{uLrelations},
we get the following equation for the flavor spinor
\beq
\label{eigeneqA}
\left[
  (\omega - \lambda\kappa ) - n^{(\lambda)}\cdot V(\omega,\vec\kappa)
\right]
\xi^{(\lambda)} = 0\,,
\eeq
where we have used the fact that
\beq
n^{(\lambda)}\cdot V = V^{(u)} - \lambda V^{(t)}\,,
\eeq
and we have indicated explicitly that, in general, $V^{(u)},V^{(t)}$ are
functions of $\omega$ and $\vec\kappa$\cite{footnote2}.

\Eq{eigeneqA} is an implicit equation that in principle determines
the dispersion relations for $\omega^{(\lambda)}(\vec\kappa)$ for each mode.
The next step is to linearize the equation, by substituting the
zeroth order solution, $\omega = \lambda\kappa$ in $V^{(u)},V^{(t)}$. Thus,
\Eq{eigeneqA} becomes
\beq
\label{eigeneqB}
H^{(\lambda)}(\vec\kappa)\,\xi^{(\lambda)}(\vec\kappa) =
\omega\,\xi^{(\lambda)}(\vec\kappa)\,,
\eeq
where
\beq
H^{(\lambda)}(\vec\kappa) = \lambda\kappa +
n^{(\lambda)}\cdot V(\lambda\kappa,\vec\kappa)\,.
\eeq
\Eq{eigeneqB} determines the positive and negative frequency dispersion
relations $\omega^{(\pm)}(\vec\kappa)$.
According to the decomposition in \Eq{Vri}, we define
\beq
H^{(\lambda)}(\vec\kappa) = H^{(\lambda)}_r(\vec\kappa) -
\frac{i}{2}\Gamma^{(\lambda)}(\vec\kappa)\,,
\eeq
where
\beqa
H^{(\lambda)}_r(\vec\kappa) & = & \lambda\kappa +
n^{(\lambda)}\cdot V_r(\lambda\kappa,\vec\kappa)\,,\nonumber\\
-\frac{1}{2}\Gamma^{(\lambda)}(\vec\kappa) & = &
n^{(\lambda)}\cdot V_i(\lambda\kappa,\vec\kappa)\,.
\eeqa
The neutrino Hamiltonian is identified, as usual, by associating it
with the positive frequency solution; that is we set
$\xi^{(\nu)}(\vec\kappa) =\xi^{(+)}(\vec\kappa)$ and
$\omega^{(\nu)}(\vec\kappa) = \omega^{(+)}(\vec\kappa)$.
For the antineutrino, we look at the equation for
$\xi^{(\bar\nu)}(\vec\kappa) \equiv (\xi^{(-)}(-\vec\kappa))^\ast$,
with the identification
$\omega^{(\bar\nu)}(\vec\kappa) = -(\omega^{(-)}(-\vec\kappa))^\ast$.
In this way, the equations are
\beq
H^{(\nu,\bar\nu)}(\vec\kappa)\,\xi^{(\nu,\bar\nu)}(\vec\kappa) =
\omega^{(\nu,\bar\nu)}(\vec\kappa)\,\xi^{(\nu,\bar\nu)}(\vec\kappa)\,,
\eeq
with
\beqa
H^{(\nu)}(\vec\kappa) & = & H^{(+)}_r(\vec\kappa) -
\frac{i}{2}\Gamma^{(+)}_i(\vec\kappa)\,,\nonumber\\
H^{(\bar\nu)}(\vec\kappa) & = &
-(H^{(-)}_r(-\vec\kappa))^\ast -
\frac{i}{2}(\Gamma^{(-)}_i(-\vec\kappa))^\ast\,,
\eeqa
or explicitly,
\beqa
H^{(\nu)}(\vec\kappa) & = & \kappa + n\cdot V_r(\kappa,\vec\kappa) +
i\,n\cdot V_i(\kappa,\vec\kappa)\,,\nonumber\\
H^{(\bar\nu)}(\vec\kappa) & = & \kappa -
n\cdot V^\ast_r(-\kappa,-\vec\kappa) +
i\,n\cdot V^\ast_i(-\kappa,-\vec\kappa)\,,
\eeqa
where $n^\mu$ has been defined in \Eq{nmu}.

In summary, for either neutrinos or antineutrinos,
we have the \emph{eigenvalue} equation for $\xi$,
\beq
\left(H_r - i\frac{\Gamma}{2}\right)\xi = \omega\xi\,,
\eeq
with $H_r$ and $\Gamma$ being Hermitian matrices in flavor space,
calculated in terms of the vector $V_\mu$,
\beqa
\label{HrGammaVrelation}
H_r & = & \left\{\begin{array}{ll}
\kappa + n\cdot V_r(\kappa,\vec\kappa) & (\nu)\\
\kappa - n\cdot V^\ast_r(-\kappa,-\vec\kappa) & (\bar\nu)
\end{array}\right.\nonumber\\[12pt]
-\frac{1}{2}\Gamma & = & \left\{\begin{array}{ll}
n\cdot V_i(\kappa,\vec\kappa) & (\nu)\\
n\cdot V^\ast_i(-\kappa,-\vec\kappa) & (\bar\nu)
\end{array}\right.\,.
\eeqa
In coordinate space, this translates to the evolution equation
\beq
\label{eveqt}
i\partial_t\xi(t) = \left(H_r - i\frac{\Gamma}{2}\right)\xi(t)\,.
\eeq
\section{Non-forward scattering terms}
\label{sec:sigma12}

\begin{figure}
\begin{center}
\epsfig{file=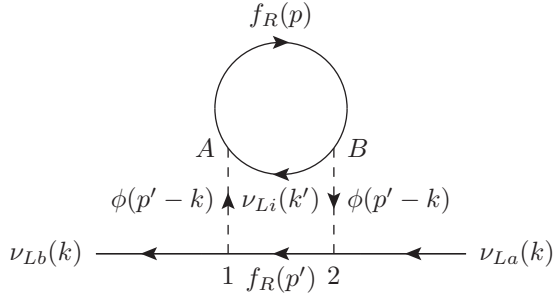,bbllx=204,bblly=356,bburx=408,bbury=466}
\end{center}
\caption{
  Two-loop diagram for the neutrino thermal self-energy matrix
  element $\Sigma_{12}$ in an $f\phi$ background.
  In principle we have to consider the various thermal vertices $A = 1,2$
  and $B = 1,2$. However, in the heavy $\phi$ limit, only the
  diagonal components of the $\phi$ thermal propagator are non-zero
  and therefore only one diagram, with $A = 1$ and $B = 2$, must be considered.
  In the labels referring to the various neutrino families,
  we use the indices $a,b$ running over the neutrino flavors
  and $i,j,k$ running over the neutrino modes with a definite dispersion
  relation in the medium. For simplicity of notation, we have
  labeled $k^\prime = p - p^\prime + k$.
  \label{fig:twoloop}
}
\end{figure}

In the case that several neutrino flavors couple to $f\phi$ as
indicated in \Eq{Lfnuphia}, the damping matrix $\Gamma$ receives
another contribution, from the diagrams depicted in \Fig{fig:twoloop}.
From a physical point of view these diagrams correspond to
contributions to the damping matrix $\Gamma$ due to the various
neutrino non-forward scattering processes that can occur in the presence
of the background particles $f\phi$, schematically
of the form $\nu_a + x \leftrightarrow \nu_b + x$, where $x = f,\phi$ (and
similar ones with $f$ and/or $\phi$ crossed). This contrasts with the
processes involved in the diagrams of \Fig{fig:oneloop}, which are
associated with decay type process like
$\nu_a + \phi \leftrightarrow f$ and related ones. To distinguish
the two types of contributions to $\Gamma$, we denote
by $\Gamma^{(1,2)}$ the contribution the one-loop diagram (\Fig{fig:oneloop})
and the two-loop diagram (\Fig{fig:twoloop}), respectively.
Our main observation is that $\Gamma^{(2)}$
has a structure that lends itself to a formulation as decoherence terms
that in turn allows us to go beyond the evolution equation \Eq{eveqt} to
consider its effects. However, before
going in that route we calculate explicitly $\Gamma^{(2)}$ 
by direct evaluation of the diagrams in \Fig{fig:twoloop}.

\subsection{Calculation of $\Sigma_{12}$ from \Fig{fig:twoloop}}

From \Fig{fig:twoloop}, taking into account
the sign difference between type 1 and type 2 vertices,
\beqa
-i\left(\Sigma_{12}(k)\right)_{ba} & = &
(-ig_a)(ig^\ast_b)\int\frac{d^4p^\prime}{(2\pi)^4}\sum_{A,B}
i\Delta^{(\phi)}_{2B}(p^\prime - k) i\Delta^{(\phi)}_{A1}(p^\prime - k)
RiS^{(f)}_{12}(p^\prime)L
\nonumber\\
&&\mbox{}\times\sum_{cd}
(ig_c)(-ig^\ast_d) (-1)\int\frac{d^4p}{(2\pi)^4}
\Tr{\left(RiS^{(f)}_{BA}(p)L (iS^{(\nu)}_{AB}(k^\prime))_{cd}\right)}\,,
\eeqa
where
\beq
\label{kprime}
k^\prime = p - p^\prime + k\,.
\eeq
Recall that $(\Sigma_i)_{ba}$ is given by \Eq{Sigmaidef1} and then
$(\Gamma^{(2)})_{ba}$ is obtained from  \Eq{HrGammaVrelation}
with $(V^\mu_i)_{ba}$ identified according to \Eq{SigmariVri}.

\begin{figure}
\begin{center}
\epsfig{file=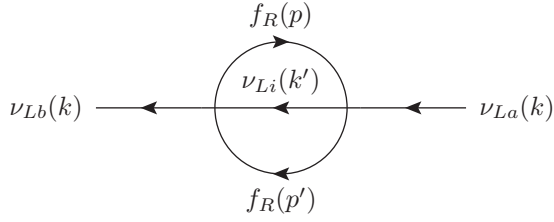,bbllx=204,bblly=361,bburx=408,bbury=441}
\end{center}
\caption{
  Collapsed version of the two-loop diagram shown in \Fig{fig:twoloop}
  in the heavy $\phi$ limit. The various labels and indices
  are the same as those in \Fig{fig:twoloop}.
  \label{fig:collapsed}
}
\end{figure}

We will assume that $m_\phi$ is larger than both the background temperature
and the incoming neutrino energy so that we can work in the
\emph{heavy $\phi$ limit}. In this case only the diagonal elements
of the thermal $\phi$ propagator are non-zero, and therefore in the
vertices in \Fig{fig:twoloop} only the case $A = 1, B = 2$
has to be considered. Then using
$\Delta^{(\phi)}_{22} = -\Delta^{(\phi)}_{11} = 1/m^2_\phi$,
\beq
\label{Sigma12A}
-i\left(\Sigma_{12}(k)\right)_{ba} = 
-\frac{g_a g^\ast_b}{m^4_\phi}\sum_{cd} g_c g^\ast_d
\int\frac{d^4p^\prime}{(2\pi)^4}RiS^{(f)}_{12}(p^\prime)L
\int\frac{d^4p}{(2\pi)^4}
\Tr{\left(RiS^{(f)}_{21}(p)L (iS^{(\nu)}_{12}(k^\prime))_{cd}\right)}\,,
\eeq
which corresponds to the collapsed diagram shown in \Fig{fig:collapsed}.
The components of the propagator matrices are given by
\beqa
S^{(f)}_{21}(p) & = & 2\pi i\delta(p^2 - m^2_f)
\left[\eta_F(p,\alpha_f) - \theta(p\cdot u)\right]\sigma_f(p)\,,\nonumber\\
S^{(f)}_{12}(p^\prime) & = & 2\pi i\delta(p^{\prime\,2} - m^2_f)
\left[\eta_F(p^\prime,\alpha_f) - \theta(-p^\prime\cdot u)\right]
\sigma_f(p^\prime)\,,
\eeqa
where
\beq
\label{sigmaf}
\sigma_f(q) = \lslash{q} + m_f\,,
\eeq
while
\beqa
\eta_F(p,\alpha_f) & = & \theta(p\cdot u)n_F(x_f) +
\theta(-p\cdot u)n_F(-x_f)\,,\nonumber\\
\eta_F(p^\prime,\alpha_f) & = & \theta(p^\prime\cdot u)n_F(x^\prime_f) +
\theta(-p^\prime\cdot u)n_F(-x^\prime_f)\,,
\eeqa
with $n_F$ defined in \Eq{nzfermion} and $\theta$ is the step function.
We have defined
\beqa
x_f & = & \beta p\cdot u - \alpha_f\,,\nonumber\\
x^\prime_f & = & \beta p^\prime\cdot u - \alpha_f\,.
\eeqa
For our purposes, it will be more convenient to use the identity
\beq
\label{nFidentity1}
n_F(-x) = 1 - n_F(x) = e^x n_F(x)\,,
\eeq
and express $S^{(f)}_{21}(p)$ and $S^{(f)}_{12}(p^\prime)$ in the form
\beqa
\label{fpropagators}
S^{(f)}_{21}(p) & = & -2\pi i\delta(p^2 - m^2_f)\sigma_f(p)
e^{x_f} n_F(x_f)\epsilon(p\cdot u)\,,\nonumber\\
S^{(f)}_{12}(p^\prime) & = & 2\pi i\delta(p^{\prime\,2} - m^2_f)
\sigma_f(p^\prime)n_F(x^\prime_f)\epsilon(p^\prime\cdot u)\,,
\eeqa
where $\epsilon(z) = \theta(z) - \theta(-z)$.

We now consider the propagator to use for the internal neutrino line.
In principle we would use the formulas appropriate for the propagating
neutrino including the background effects, taking into account the
relationship between the neutrino flavor states and the mode
states with definite dispersion relation. However, we adopt the
perturbative approach and neglect the effect of the non-zero neutrino
masses and/or dispersion relations in the calculation of $\Sigma_{12}$.
In this case the neutrino propagator $S^{(\nu)}_{AB}(k^\prime))_{cd}$
is diagonal in flavor space, with all the elements actually being the
same since all the neutrinos have the same mass (zero) and the same
chemical potential. Specifically,
\beq
\label{nupropagator}
(S^{(\nu)}_{12}(k^\prime))_{cd} = 2\pi i\delta(k^{\prime\,2})
\sigma_\nu(k^\prime)n_F(x^\prime_\nu)\epsilon(k^\prime\cdot u)\delta_{cd}\,,
\eeq
where
\beq
\label{xnu}
x^\prime_\nu = \beta k^\prime\cdot u - \alpha_\nu\,,
\eeq
and
\beq
\label{sigmanu}
\sigma^{(\nu)}(k^\prime) = L\lslash{k}^\prime\,.
\eeq

We now work the product of the fermion propagators in \Eq{Sigma12A}
as follows. Using the relation
\beq
x_\nu + x_f = x^\prime_\nu + x^\prime_f\,,
\eeq
which follows from \Eq{kprime}, the following identity is readily derived
(see \Appendix{sec:nidentities}),
\beq
\label{nidentities}
\frac{1}{n_F(x_\nu)}e^{x_f} n_F(x_f) n_F(x^\prime_f) n_F(x^\prime_\nu) = E\,,
\eeq
where
\beq
\label{E}
E \equiv n_F(x_f)(1 - n_F(x^\prime_f)) -
n_F(x^\prime_\nu)(n_F(x_f) - n_F(x^\prime_f))\,.
\eeq
Substituting \Eqs{fpropagators}{nupropagator} in \Eq{Sigma12A},
and using \Eq{nidentities}, from \Eq{Sigmaidef1} we then obtain
\beqa
\left(\Sigma_i(k)\right)_{ba} & = & 
-K_{ba}\int\frac{d^4p^\prime}{(2\pi)^3}\int\frac{d^4p}{(2\pi)^3}
(2\pi)\delta(p^{\prime\,2} - m^2_f)\delta(p^2 - m^2_f)\delta(k^{\prime\,2})
\nonumber\\
&&\mbox{}\times 
\epsilon(p\cdot u)\epsilon(p^\prime\cdot u)\epsilon(k^\prime\cdot u)
R\sigma_f(p^\prime)L\,
\Tr\{R\sigma_f(p)L\sigma_{\nu}(k^\prime)\}E\,,
\eeqa
where
\beq
\label{K}
K_{ba} = \frac{g_a g^\ast_b}{2m^4_\phi}\left(\sum_{c}|g_c|^2\right)\,.
\eeq
We now let $k^\prime$ be an arbitrary variable, but insert the factor
$\delta^{(4)}(k^\prime + p^\prime - p - k)$ and integrate over $k^\prime$.
Thus,
\beqa
\label{Sigmaiba}
\left(\Sigma_i(k)\right)_{ba} & = & 
-K_{ba}\int\frac{d^4p^\prime}{(2\pi)^3}
\frac{d^4p}{(2\pi)^3}
\frac{d^4k^\prime}{(2\pi)^3}
\delta(p^{\prime\,2} - m^2_f)\delta(p^2 - m^2_f)\delta(k^{\prime\,2})
\epsilon(p\cdot u)\epsilon(p^\prime\cdot u)\epsilon(k^\prime\cdot u)
\nonumber\\
&&\mbox{}\times (2\pi)^4\delta^{(4)}(k^\prime + p^\prime - p - k)
R\sigma_f(p^\prime)L\,
\Tr\{R\sigma_f(p)L\sigma_{\nu}(k^\prime)\}E\,.
\eeqa

The next step is to carry out the integrals over
$p^0, p^{\prime\,0}, k^{\prime\,0}$. We do
the integral over $k^{\prime\,0}$ first.
With the help of the delta function we obtain
\beqa
\label{Sigma12B1}
\left(\Sigma_i(k)\right)_{ba} & = &
-K_{ba}\int\frac{d^4p^\prime}{(2\pi)^3}
\frac{d^4p}{(2\pi)^3}
\frac{d^3\kappa^\prime}{(2\pi)^3 2\omega_{\kappa^\prime}}
\delta(p^{\prime\,2} - m^2_f)\delta(p^2 - m^2_f)
\epsilon(p\cdot u)\epsilon(p^\prime\cdot u)
\nonumber\\
&& \times (2\pi)^4\left\{
\delta^{(4)}(k + p - k^\prime - p^\prime)
R\sigma_f(p^\prime)L\,
\Tr\{R\sigma_f(p)L\sigma_\nu(k^\prime)\}E_\nu\right.
\nonumber\\
&& - \left.\delta^{(4)}(k + p + k^\prime - p^\prime)
R\sigma_f(p^\prime)L\,
\Tr\{R\sigma_f(p)L\sigma_{\nu}(-k^\prime)\}E_{\bar\nu}
\right\}\,,
\eeqa
where
\beqa
\label{Enu}
E_\nu & = & n_F(x_f)(1 - n_F(x^\prime_f)) -
f_\nu(\omega_{\kappa^\prime})(n_F(x_f) - n_F(x^\prime_f))
\,,\nonumber\\
E_{\bar\nu} & = & n_F(x^\prime_f)(1 - n_F(x_f)) +
f_{\bar\nu}(\omega_{\kappa^\prime})(n_F(x_f) - n_F(x^\prime_f))
\,,
\eeqa
with
\beqa
k^{\prime\,\mu} & = & (\omega_{\kappa^\prime},\vec\kappa^\prime)
\,,\nonumber\\
\omega_{\kappa^\prime} & = & |\vec\kappa^\prime|\,.
\eeqa
To arrive at \Eq{Sigma12B1} we have changed variables
$\vec\kappa \rightarrow -\vec\kappa$ in the term with the 
$\sigma_{\nu}(-k^\prime)$ factor, and we have defined
\beqa
\label{Enudef}
E_\nu & \equiv &
\left.E\right|_{\omega^\prime = \omega_{\kappa^\prime}}\,,
\nonumber\\
E_{\bar\nu} & \equiv &
\left.E\right|_{\omega^\prime = -\omega_{\kappa^\prime}}\,,
\eeqa
with $E$ defined in \Eq{E}. Using the relations
\beqa
\label{nfnurelations}
\left.n_F(x^\prime_\nu)\right|_{
  \omega^\prime = \omega_{\kappa^\prime}
} & = & f_\nu(\omega_{\kappa^\prime})\,,\nonumber\\
\left.n_F(x^\prime_\nu)\right|_{
  \omega^\prime = -\omega_{\kappa^\prime}
} & = & 1 - f_{\bar\nu}(\omega_{\kappa^\prime})\,,
\eeqa
$E_{\nu,\bar\nu}$ reduce to the formulas given in \Eq{Enu}.

Next we carry out the integration over $p^0, p^{\prime\,0}$ in a similar way.
In analogy with \Eq{nfnurelations}, we will use
\beqa
\label{nfrelations}
\left.n_F(x_f)\right|_{p^0 = E_p} & = & f_f(E_p)\,,\nonumber\\
\left.n_F(x_f)\right|_{p^0 = -E_p} & = & 1 - f_{\bar f}(E_p)\,,
\eeqa
with
\beq
E_p = \sqrt{{\vec p}^{\;2} + m^2_f}\,.
\eeq
and the corresponding formulas for $p^\prime$. Thus, from \Eq{Sigma12B1}
we obtain
\beqa
\label{Sigma12B2}
\left(\Sigma_i(k)\right)_{ba} & = & 
-K_{ba}\int\frac{d^3p^\prime}{(2\pi)^3 2E_{p^\prime}}
\frac{d^3p}{(2\pi)^3 2E_{p}}
\frac{d^3\kappa^\prime}{(2\pi)^3 2\omega_{\kappa^\prime}}
\nonumber\\
&& \times (2\pi)^4\left\{
\delta^{(4)}(k + p - k^\prime - p^\prime)
R\sigma_f(p^\prime)L\,
\Tr\{R\sigma_f(p)L\sigma_\nu(k^\prime)\}E_{\nu,++}
\right.\nonumber\\
&& - \left.\delta^{(4)}(k - p - k^\prime - p^\prime)
R\sigma_f(p^\prime)L\,
\Tr\{R\sigma_f(-p)L\sigma_\nu(k^\prime)\}E_{\nu,-+}
\right.\nonumber\\
&& - \left.\delta^{(4)}(k + p + p^\prime - k^\prime)
R\sigma_f(-p^\prime)L\,
\Tr\{R\sigma_f(p)L\sigma_\nu(k^\prime)\}E_{\nu,+-}
\right.\nonumber\\
&& + \left.\delta^{(4)}(k + p^\prime - k^\prime - p)
R\sigma_f(-p^\prime)L\,
\Tr\{R\sigma_f(-p)L\sigma_\nu(k^\prime)\}E_{\nu,--}
\right.\nonumber\\
&& - \left.\delta^{(4)}(k + p + k^\prime - p^\prime)
R\sigma_f(p^\prime)L\,
\Tr\{R\sigma_f(p)L\sigma_\nu(-k^\prime)\}E_{\bar\nu,++}
\right.\nonumber\\
&& + \left.\delta^{(4)}(k + k^\prime - p^\prime - p)
R\sigma_f(p^\prime)L\,
\Tr\{R\sigma_f(-p)L\sigma_\nu(-k^\prime)\}
E_{\bar\nu,-+}\right.\nonumber\\
&& + \left.\delta^{(4)}(k + p + p^\prime + k^\prime)
R\sigma_f(-p^\prime)L\,
\Tr\{R\sigma_f(p)L\sigma_\nu(-k^\prime)\}
E_{\bar\nu,+-}\right.\nonumber\\
&& - \left.\delta^{(4)}(k + p^\prime + k^\prime - p)
R\sigma_f(-p^\prime)L\,
\Tr\{R\sigma_f(-p)L\sigma_\nu(-k^\prime)\}
E_{\bar\nu,--}\right\}\,,
\eeqa
with
\beq
p^\mu = (E_p,\vec p)\,,
\eeq
and similarly for $p^{\prime\,\mu}$. In \Eq{Sigma12B2} we have introduced
the factors $E_{\nu,\lambda\lambda^\prime}$ and
$E_{\bar\nu,\lambda\lambda^\prime}$
(with $\lambda,\lambda^\prime$ being $\pm$), which are defined as follows,
\beqa
E_{\nu,\lambda\lambda^\prime} = \left.E_\nu\right|_{
p^0 = \lambda E_p,\,p^{\prime\,0} = \lambda^\prime E_{p^\prime}}
\eeqa
and similarly for $E_{\bar\nu,\lambda\lambda^\prime}$.
Using \Eq{nfrelations} and the corresponding formulas
for $n_F(x^\prime_f)$ in \Eq{Enu}, the explicit
formulas are given in \Table{table:processes}.
%
%
%
\begin{table}
\begin{center}
  \begin{tabular}{|l|l|}
    \hline
  $E_{\nu,++} = \fEp(1 - \fEpp) - \fnu(\fEp - \fEpp)$ &
  $\nu_{a,b}(k) + f(p) \leftrightarrow \nu_i(k^\prime) + f(p^\prime)$\\
  $E_{\nu,-+} = (1 - \fbEp)(1 - \fEpp) - \fnu(1 - \fbEp - \fEpp)$ &
  $\nu_{a,b}(k) \leftrightarrow \nu_i(k^\prime) + \bar f(p) + f(p^\prime)$\\
  $E_{\nu,+-} = \fEp\fbEpp - \fnu(\fEp + \fbEpp - 1)$ &
  $\nu_{a,b}(k) + f(p) + \bar f(p^\prime) \leftrightarrow \nu_i(k^\prime)$\\
  $E_{\nu,--} = (1 - \fbEp)\fbEpp - \fnu(\fbEpp - \fbEp)$ &    
  $\nu_{a,b}(k) + \bar f(p^\prime) \leftrightarrow
  \nu_i(k^\prime) + \bar f(p)$\\
  $E_{\bar\nu,++} = (1 - \fEp)\fEpp + \fnub(\fEp - \fEpp)$ &
  $\nu_{a,b}(k) + \bar\nu_i(\bar k^\prime) + f(p)\leftrightarrow f(p^\prime)$\\
  $E_{\bar\nu,-+} = \fbEp\fEpp + \fnub(1 - \fbEp - \fEpp)$ &
  $\nu_{a,b}(k) + \bar\nu_i(\bar k^\prime) \leftrightarrow
  \bar f(p) + f(p^\prime)$\\
  $E_{\bar\nu,+-} = (1 - \fEp)(1 - \fbEpp) + \fnub(\fEp + \fbEpp - 1)$ &
  $\nu_{a,b}(k) + \bar \nu_i(\bar k^\prime) +
  f(p) + \bar f(p^\prime) \leftrightarrow 0$\\
  $E_{\bar\nu,--} = \fbEp(1 - \fbEpp) + \fnub(\fbEpp - \fbEp)$ &
  $\nu_{a,b}(k) + \bar\nu_i(\bar k^\prime) + \bar f(p^\prime) \leftrightarrow
  \bar f(p)$\\
  \hline
\end{tabular}
\caption{Correspondence between the $E_{\nu,\lambda\lambda^\prime}$
  and $E_{\bar\nu,\lambda\lambda^\prime}$ factors defined in \Eq{Enu},
  and the process that contributes to the $\nu(k)$ damping via
  \Eq{Sigma12B2}. To simplify the notation we are using
    the shorthands shown in \Eq{fshorthand} for the various distribution
    functions.
  \label{table:processes}
}
\end{center}
\end{table}
To simplify the notation in the formulas summarized in \Table{table:processes}
we have introduce the shorthands
\beqa
\label{fshorthand}
f = f_f(E_p), \quad f^\prime = f_f(E_{p^\prime}), \quad
f^\prime_\nu = {f_\nu(\omega_{\kappa^\prime})}\nonumber\\
\bar f = f_{\bar f}(E_p), \quad \bar f^\prime = f_{\bar f}(E_{p^\prime}), \quad
\bar f^\prime_\nu = f_{\bar\nu}(\omega_{\kappa^\prime})\,.
\eeqa
The formulas for $E_{\bar\nu,\lambda\lambda^\prime}$ are obtained from
those for $E_{\nu,\lambda\lambda^\prime}$ by making the replacement
$\fnu \rightarrow (1 - \fnub)$.

Using the relations
\beqa
\sigma_f(p) & = & \sum_s u_f(\vec p,s)\bar u_f(\vec p,s)\,,\nonumber\\
\sigma_f(-p) & = & -\sum_s v_f(\vec p,s)\bar v_f(\vec p,s)\,,
\eeqa
and the analogous relations for $\sigma_\nu(k^\prime)$,
makes it evident that the matrix element
\beq
\bar u_{\nu_b}(\vec\kappa) \left(\Sigma_i(k)\right)_{ba} u_{\nu_a}(\vec\kappa)
\eeq
can be expressed as a sum of terms of the form
\beq
M\left(\nu_a(k) + f(p) \rightarrow \nu_i(k^\prime) + f(p^\prime)\right)
M^\ast\left(\nu_b(k) + f(p) \rightarrow \nu_i(k^\prime) + f(p^\prime)\right)\,,
\eeq
involving the amplitudes for the processes
\beq
\label{process1}
\nu_{a,b}(k) + f(p) \leftrightarrow \nu_i(k^\prime) + f(p^\prime)\,,
\eeq
as well as the processes obtained by crossing
$f(p), f(p^\prime),\nu_i(k^\prime)$.
Each of the terms that appear within the
bracket in \Eq{Sigma12B2} corresponds to one such process, and its inverse.
The factors of $E_{\nu,\lambda\lambda^\prime},E_{\bar\nu,\lambda\lambda^\prime}$
incorporate the statistical effects of the background.
As is well known\cite{Kadanoff:1962}, for Fermi systems the inverse
reactions are inhibited as a consequence of the Pauli blocking effect,
and they contribute additively to the depletion of the
state. The formulas given in \Table{table:processes} reflect the
fact that the contributions from the direct and the inverse process
are given by the sum of their rates instead of
their difference as in the bosonic case. For example,
$E_{\nu,++}$ can be rewritten in the form
\beq
E_{\nu,++} = \fEp(1 - \fEpp)(1 - \fnu) + \fnu\fEpp(1 - \fEp)\,,
\eeq
which is just the sum of the statistical factors for the direct and inverse
process indicated in \Eq{process1}. In similar fashion
it can be verified that the terms in \Eq{Sigma12B2} and
the associated $E_{\nu,\lambda\lambda^\prime}$
and $E_{\bar\nu,\lambda\lambda^\prime}$ can be identified with the
various processes as indicated in \Table{table:processes}.
Similar identifications can be made for the antineutrino matrix element
\beq
\bar v_{\nu_b}(\vec\kappa) \left(\Sigma_i(-k)\right)_{ba}
v_{\nu_a}(\vec\kappa)\,.
\eeq

For some conditions, some of these processes will be kinematically
forbidden and will not contribute.
We now assume that the conditions are such that, for $\omega > 0$,
the only processes that are
kinematically accessible are the one shown above, and the following one,
\beq
\nu_{a,b}(k) + \bar f(p^\prime) \rightarrow \nu_i(k^\prime) + \bar f(p)\,.
\eeq
These correspond to the the first and the fourth
terms, respectively, in the list of terms that appear within the
bracket in \Eq{Sigma12B2}. Alternatively, for $\omega < 0$, the only
kinematically accessible processes are
\beqa
\bar\nu_{a,b}(k) + f(p^\prime) & \rightarrow &
\bar \nu_i(k^\prime) + f(p)\,,\nonumber\\
\bar\nu_{a,b}(k) + \bar f(p) & \rightarrow &
\bar\nu_i(k^\prime) + \bar f(p^\prime)\,,
\eeqa
which correspond to the fifth and eighth terms within the
bracket in \Eq{Sigma12B2}. In addition we will assume that there
no neutrinos or antineutrinos in the background, therefore we set
$f_\nu$ and $f_{\bar\nu}$ to zero. Then,
\beqa
\label{Sigma12C}
\left(\Sigma_i(k)\right)_{ba} & = & 
-K_{ba}\int\frac{d^3p^\prime}{(2\pi)^3 2E_{p^\prime}}
\frac{d^3p}{(2\pi)^3 2E_{p}}
\frac{d^3\kappa^\prime}{(2\pi)^3 2\omega_{\kappa^\prime}}
\nonumber\\
&& \times  (2\pi)^4\left\{
\delta^{(4)}(k + p - k^\prime - p^\prime)
R\sigma_f(p^\prime)L\,
\Tr\{R\sigma_f(p)L\sigma_{\nu}(k^\prime)\}
\left[f_{f}(E_p)(1 - f_{f}(E_{p^\prime}))\right]\right.\nonumber\\
&& + \left.\delta^{(4)}(k + p^\prime - k^\prime - p)
R\sigma(-p^\prime)L\,
\Tr\{R\sigma(-p)L\sigma_{\nu}(k^\prime)\}
\left[(1 - f_{\bar f}(E_p))f_{\bar f}(E_{p^\prime})\right]
\right\}\,.\nonumber\\
&& - \left.\delta^{(4)}(k + p + k^\prime - p^\prime)
R\sigma_f(p^\prime)L\,
\Tr\{R\sigma_f(p)L\sigma_{\nu}(-k^\prime)\}
\left[(1 - f_{f}(E_{p}))f_{f}(E_{p^\prime})\right]\right.\nonumber\\
&& - \left.\delta^{(4)}(k + p^\prime + k^\prime - p)
R\sigma_f(-p^\prime)L\,
\Tr\{R\sigma_f(-p)L\sigma_{\nu}(-k^\prime)\}
\left[f_{\bar f}(E_{p})(1 - f_{\bar f}(E_{p^\prime}))\right]
\right\}\,,\nonumber\\
\eeqa
where, as we have mentioned, if $\omega > 0$ only the first two
terms in the bracket contribute, while for $\omega < 0$ only the last
two contribute.

Remembering \Eq{SigmaV}, we identify
$\left(V^\mu_i(\omega,\vec\kappa)\right)_{ba}$ by writing
\beq
\label{SigmaiVi}
\left(\Sigma_i(k)\right)_{ba} =
\left(V^\mu_i(\omega,\vec\kappa)\right)_{ba} \gamma_\mu L\,.
\eeq

\subsection{Formula for $V^\mu_i$}
  
We now express $V^\mu_i$ as follows. Using \Eqs{sigmaf}{sigmanu},
\beq
\Tr\{R\sigma_f(p)L\sigma_{\nu}(k^\prime)\} = 
-\Tr\{R\sigma_f(p)L\sigma_{\nu}(-k^\prime)\} = 2p\cdot k^\prime\,,
\eeq
and we have
\beq
\label{ViA1}
(V^\mu_i)_{ba} = (V^{(+)\mu}_i)_{ba} + (V^{(-)\mu}_i)_{ba}\,,
\eeq
where
\beqa
\label{ViA2}
(V^{(+)\mu}_i)_{ba} & = & 
-2K_{ba}\int\frac{d^3\kappa^\prime}{(2\pi)^3 2\omega_{\kappa^\prime}}
I^{(+)\mu\nu}(k - k^\prime)k^\prime_\nu\nonumber\\
(V^{(-)\mu}_i)_{ba} & = &
-2K_{ba}\int\frac{d^3\kappa^\prime}{(2\pi)^3 2\omega_{\kappa^\prime}}
I^{(-)\mu\nu}(k + k^\prime)k^\prime_\nu\,,
\eeqa
with
\beqa
\label{Imunuq}
I^{(+)}_{\mu\nu}(q) & = & 
\int\frac{d^3p^\prime}{(2\pi)^3 2E_{p^\prime}}
\frac{d^3p}{(2\pi)^3 2E_{p}}\nonumber\\
&& \times (2\pi)^4\left\{
\delta^{(4)}(p + q - p^\prime)
f_{f}(E_{p})\left(1 - f_{f}(E_{p^\prime})\right)\right.\nonumber\\
&& + \left.\delta^{(4)}(p^\prime + q - p)
\left(1 - f_{\bar f}(E_{p})\right)f_{\bar f}(E_{p^\prime})
\right\}p^\prime_\mu p_\nu\,,\nonumber\\
I^{(-)}_{\mu\nu}(q) & = & 
\int\frac{d^3p^\prime}{(2\pi)^3 2E_{p^\prime}}
\frac{d^3p}{(2\pi)^3 2E_{p}}\nonumber\\
&& \times (2\pi)^4\left\{
\delta^{(4)}(p + q - p^\prime)
\left(1 - f_{f}(E_{p})f_{f}(E_{p^\prime})\right)\right.\nonumber\\
&& + \left.\delta^{(4)}(p^\prime + q - p)
f_{\bar f}(E_{p})\left(1 - f_{\bar f}(E_{p^\prime})\right)
\right\}p^\prime_\mu p_\nu\,.
\eeqa
It is useful to note that
\beq
\label{I+-relation}
I^{(-)}_{\mu\nu}(q) = I^{(+)}_{\nu\mu}(-q)\,.
\eeq
The remark below \Eq{Sigma12C} is equivalent to say that
$V^{(-)\mu}_i$ or $ V^{(+)\mu}_i$ is zero if $\omega > 0$
or $\omega < 0$, respectively, that is
\beqa
\label{Vpmnonzero}
V^{(-)\mu}_i & = & 0 \qquad \mbox{if } \omega > 0\,,\nonumber\\
V^{(+)\mu}_i & = & 0 \qquad \mbox{if } \omega < 0\,.
\eeqa
Thus, finally, putting
\beq
\label{kprimenprimerel}
k^{\prime\,\mu} = \omega_{\kappa^\prime} n^{\prime\,\mu}\,,
\eeq
with
\beq
n^{\prime\,\mu} = (1,\hat\kappa^\prime)\,,
\eeq
we have
\beqa
\label{ViB3}
\left(V^{(+)\mu}_i(\omega,\vec\kappa)\right)_{ba} & = &
-\frac{g_a g^\ast_b}{2m^4_\phi}\left(\sum_{c} |g_c|^2\right)
\int\frac{d^3\kappa^\prime}{(2\pi)^3}
I^{(+)\mu\nu}(k - k^\prime)n^{\prime}_{\nu}\,,\nonumber\\
\left(V^{(-)\mu}_i(\omega,\vec\kappa)\right)_{ba} & = &
-\frac{g_a g^\ast_b}{2m^4_\phi}\left(\sum_{c} |g_c|^2\right)
\int\frac{d^3\kappa^\prime}{(2\pi)^3}
I^{(-)\mu\nu}(k + k^\prime)n^{\prime}_{\nu}\,,
\eeqa
where we have substituted the explicit expression for $K_{ba}$ defined
in \Eq{K}.

\subsection{Formula for $\Gamma^{(2)}$}

From \Eq{HrGammaVrelation}, remembering \Eq{Vpmnonzero},
\beq
\label{Gammafinalformula}
-\frac{1}{2}\Gamma^{(2)} = \left\{\begin{array}{ll}
n\cdot V^{(+)}_i(\kappa,\vec\kappa) & (\nu)\\
n\cdot V^{(-)\ast}_i(-\kappa,-\vec\kappa) & (\bar\nu)
\end{array}\right.\,.
\eeq
Denoting by $\Gamma^{(\nu)},\Gamma^{(\bar\nu)}$ the matrix for neutrinos or
antineutrinos, respectively, explicitly using \Eq{ViB3},
\beqa
\label{Gammaexplicit}
\Gamma^{(\nu)} & = &
g_a g^\ast_b\left(\sum_{c} |g_c|^2\right)\gamma^{(\nu)}\,,\nonumber\\
\Gamma^{(\bar\nu)} & = &
g^\ast_a g_b\left(\sum_{c} |g_c|^2\right)\gamma^{(\bar\nu)}\,,
\eeqa
with
\beqa
\label{gammalinblad}
\gamma^{(\nu)} & = & \frac{1}{m^4_\phi}
n_\mu \int\frac{d^3\kappa^\prime}{(2\pi)^3}
I^{(+)\mu\nu}(k - k^\prime) n^\prime_\nu\,,\nonumber\\
\gamma^{(\bar\nu)} & = & \frac{1}{m^4_\phi}
n_\mu \int\frac{d^3\kappa^\prime}{(2\pi)^3}
I^{(+)\nu\mu}(k - k^\prime)n^{\prime}_{\nu}\,,
\eeqa
where we have used \Eq{I+-relation}.

The integral expressions in \Eq{gammalinblad} can be simplified as follows.
The relevant integrals for neutrinos, or antineutrinos are
\beqa
\label{Inu12}
n_\mu \int\frac{d^3\kappa^\prime}{(2\pi)^3}
I^{(+)\mu\nu}(k - k^\prime)n^{\prime}_{\nu} & = &
I^{(f)}_1 + I^{(\bar f)}_2\,,\nonumber\\
n_\mu \int\frac{d^3\kappa^\prime}{(2\pi)^3}
I^{(+)\nu\mu}(k - k^\prime)n^{\prime}_{\nu} & = &
I^{(f)}_2 + I^{(\bar f)}_1\,,
\eeqa
respectively, where, for $x = f,\bar f$, we define
\beqa
\label{IIbardef}
I^{(x)}_1 & = & 
\int\frac{d^3p}{(2\pi)^3 2E_{p}}
\frac{d^3p^\prime}{(2\pi)^3 2E_{p^\prime}}
\frac{d^3\kappa^\prime}{(2\pi)^3}
\nonumber\\
&& \times (2\pi)^4
\delta^{(4)}(p + k - p^\prime - k^\prime)
f_{x}(E_{p})\left(1 - f_{x}(E_{p^\prime})\right)
(p\cdot n^\prime)(p^\prime\cdot n)\,,\nonumber\\
& = & \frac{2}{\omega_\kappa}\int\frac{d^3p}{(2\pi)^3 2E_{p}}
\frac{d^3p^\prime}{(2\pi)^3 2E_{p^\prime}}
\frac{d^3\kappa^\prime}{(2\pi)^3 2\omega_{\kappa^\prime}}
\nonumber\\
&& \times (2\pi)^4
\delta^{(4)}(p + k - p^\prime - k^\prime)
f_{x}(E_{p})\left(1 - f_{x}(E_{p^\prime})\right)
(p\cdot k^\prime)(p^\prime\cdot k)\,,\nonumber\\
I^{(x)}_2 & = & 
\int\frac{d^3p}{(2\pi)^3 2E_{p}}
\frac{d^3p^\prime}{(2\pi)^3 2E_{p^\prime}}
\frac{d^3\kappa^\prime}{(2\pi)^3}
\nonumber\\
&& \times (2\pi)^4
\delta^{(4)}(p^\prime + k - p - k^\prime)
\left(1 - f_{x}(E_{p})\right)f_{x}(E_{p^\prime})
(p\cdot n^\prime)(p^\prime\cdot n)\nonumber\\
%
%
& = & \frac{2}{\omega_{\kappa}}\int\frac{d^3p}{(2\pi)^3 2E_{p}}
\frac{d^3p^\prime}{(2\pi)^3 2E_{p^\prime}}
\frac{d^3\kappa^\prime}{(2\pi)^3 2\omega_{\kappa^\prime}}
\nonumber\\
&& \times (2\pi)^4
\delta^{(4)}(p + k - p^\prime - k^\prime)
f_{x}(E_{p})\left(1 - f_{x}(E_{p^\prime})\right)
(p^\prime\cdot k^\prime)(p\cdot k)\,,
\eeqa
We have used \Eq{kprimenprimerel} and the analogous relation between
$k^\mu$ and $n^\mu$, and in the expression for $I^{(x)}_2$ we have relabeled
$p$ and $p^\prime$.
For $I^{(x)}_1$ we use the fact that the delta function
implies that $p^\prime\cdot k = p\cdot k^\prime$, while for
$I^{(x)}_2$ we use $p\cdot k = p^\prime\cdot k^\prime$. Therefore,
\beqa
\label{Ix12final}
I^{(x)}_1 & = & \frac{2}{\omega_\kappa}\int\frac{d^3p}{(2\pi)^3 2E_{p}}
\frac{d^3p^\prime}{(2\pi)^3 2E_{p^\prime}}
\frac{d^3\kappa^\prime}{(2\pi)^3 2\omega_{\kappa^\prime}}
\nonumber\\
&& \times (2\pi)^4
\delta^{(4)}(p + k - p^\prime - k^\prime)
f_{x}(E_{p})\left(1 - f_{x}(E_{p^\prime})\right)
(p\cdot k^\prime)^2\,,\nonumber\\
I^{(x)}_2 & = & \frac{2}{\omega_\kappa}\int\frac{d^3p}{(2\pi)^3 2E_{p}}
\frac{d^3p^\prime}{(2\pi)^3 2E_{p^\prime}}
\frac{d^3\kappa^\prime}{(2\pi)^3 2\omega_{\kappa^\prime}}
\nonumber\\
&& \times (2\pi)^4
\delta^{(4)}(p + k - p^\prime - k^\prime)
f_{x}(E_{p})\left(1 - f_{x}(E_{p^\prime})\right)
(p\cdot k)^2\,.
\eeqa
From \Eqs{gammalinblad}{Inu12} we then have
\beqa
\label{gammalinbladfinal}
\gamma^{(\nu)} & = & \frac{1}{m^4_\phi}(I^{(f)}_1 + I^{(\bar f)}_2)
\,,\nonumber\\
\gamma^{(\bar\nu)} & = & \frac{1}{m^4_\phi}(I^{(f)}_2 + I^{(\bar f)}_1)\,.
\eeqa
\section{Non-forward scattering as a decoherence effect}
\label{sec:decoherence}

Our main observation here is that $\Gamma^{(2)}$, given
in \Eq{Gammaexplicit} by direct evaluation of the diagrams in \Fig{fig:twoloop},
has a structure that lends itself to a formulation as decoherence terms
in the context of the Linblad equation, and the notion of the
stochastic evolution of the state vector\cite{Daley:2014fha,Weinberg:2011jg,
pearle,Plenio:1997ep,Lieu:2019cev}).
Thus we will assume that, for kinematic reasons, $\Gamma^{(1)}$ is zero
and that $\Gamma^{(2)}$ is the only contribution to the damping matrix. 
The idea then is to assume that the evolution due to the damping effects
described by $\Gamma^{(2)}$ is accompanied by a stochastic evolution
that cannot be described by the coherent evolution of the state vector.

Let us then consider the evolution of the state vector
(using a generic notation)
\beq
i\partial_t\phi(t) = H\phi(t)\,,
\eeq
with
\beq
H = H_r - \frac{i}{2}\Gamma\,.
\eeq
In an interval $dt$ the state vector would have evolved coherently
to
\beq
\phi_1(t + dt) \equiv (1 - iHdt)\phi\,.
\eeq
The norm of this vector is
\beq
\phi^\dagger_1(t + dt)\phi_1(t + dt) = 1 - p\,,
\eeq
where
\beq
\label{pdef}
p \equiv \phi^\dagger\Gamma\phi dt\,.
\eeq
Thus, we interpret $p$ as the probability that the system has decayed
($1 - p$ is the survival probability) due to the coherent (but non-Hermitian)
evolution. We now assume that this coherent evolution is accompanied by
stochastic processes that cause the system to ``jump'' from the
initial state to a set of possible states, thus causing the damping.

To define the construction, suppose specifically that $\Gamma$ has the form
\beq
\label{GammaLi}
\Gamma = \sum_i L^\dagger_i L_i \,.
\eeq
In other words, suppose that we can find a set of matrices $L_i$ such
that $\Gamma$ can be written in this form. As we will verify later on,
this is indeed the case for $\Gamma^{(2)}$. Let us call
\beq
p_i \equiv \phi^\dagger L^\dagger_i L_i\phi\,.
\eeq
Therefore, from \Eq{pdef},
\beq
p = dt\sum_i p_i\,.
\eeq
Now we want to say that the stochastic processes cause the state vector to
``jump'' to any of the (normalized) state vectors
\beq
\phi^{(i)} \equiv \frac{L_i\phi}{\sqrt{\phi^\dagger L^\dagger_i L_i\phi}} =
\frac{L_i\phi}{\sqrt{p_i}}\,,
\eeq
with a probability $\pi_i$. Of course the condition is that
\beq
\sum_i \pi_i = p\,,
\eeq
which we satisfy by assuming that
\beq
\pi_i = p_i dt\,.
\eeq
The main assumption is that the evolution of the system, taking into account
both the coherent and stochastic evolution, is described by the density
matrix (in the sense that we can use it to calculate averages of quantum
expectation values)
\beqa
\rho(t + dt) & = & \phi_1\phi^\dagger_1 +
\sum_i \pi_i \phi^{(i)} \phi^{(i)\dagger}\nonumber\\
& = & (1 - iHdt)\rho(1 + iH^\dagger dt) + dt\sum_i L_i \rho L^\dagger_i
\nonumber\\
& = & \rho(t) - i(H\rho - \rho H^\dagger)dt + dt\sum_i L_i \rho L^\dagger_i
\nonumber\\
& = & \rho(t) - i[H_r,\rho]dt - \frac{1}{2}\{\Gamma,\rho\}dt +
dt\sum_i L_i \rho L^\dagger_i\,,
\eeqa
or
\beq
\partial_t\rho = -i[H_r,\rho] + \sum_i
\left\{L_i \rho L^\dagger_i - \frac{1}{2}L^\dagger_i L_i\rho -
\frac{1}{2}\rho L^\dagger_i L_i\right\}\,,
\eeq
which is the Linblad equation\cite{footnote3}.

From \Eq{Gammaexplicit}, it is immediately evident that $\Gamma^{(2)}$
has the form given in \Eq{GammaLi}. Indeed, to be more precise, 
only one such matrix $L$ is needed, for $\ell = \nu,\bar\nu$,
\beqa
\label{Gammalinblad}
\Gamma^{(\ell)} & = & L^{(\ell)\dagger} L^{(\ell)}\,,\nonumber\\
\Gamma^{(\ell)}_{ba} & = & \sum_c (L^{(\ell)\dagger})_{bc} L^{(\ell)}_{ca} =
\sum_c L^{(\ell)\ast}_{cb} L^{(\ell)}_{ca} \,,
\eeqa
with the identification
\beqa
\label{Lxidentification}
L^{(\nu)}_{ca} & \equiv & \sqrt{\gamma^{(\nu)}}\, g_c g_a \,,\nonumber\\
L^{(\bar\nu)}_{ca} & \equiv & \sqrt{\gamma^{(\bar\nu)}}\, g^\ast_c g^\ast_a\,.
\eeqa
In summary we assert that, in the situation that $\Gamma^{(1)}$
is zero (or negligible) so that the damping matrix is given by $\Gamma^{(2)}$,
determined from \Fig{fig:twoloop} and which has the form given in
\Eq{Gammaexplicit} (under the approximations and idealizations we have made),
then its effects are more effectively taken into account in the context of
the evolution equation for the flavor density matrix, in this case,
\beq
\partial_t\rho^{(\ell)} = -i[H^{(\ell)}_r,\rho^{(\ell)}] +
\left\{L^{(\ell)} \rho^{(\ell)} L^{(\ell)\dagger} -
\frac{1}{2}L^{(\ell)\dagger} L^{(\ell)}\rho^{(\ell)} -
\frac{1}{2}\rho^{(\ell)} L^{(\ell)\dagger} L^{(\ell)}\right\}\,,
\eeq
for neutrinos or antineutrinos, with $L^{(\ell)}$
identified in \Eq{Lxidentification}.
In what follows we compute the integrals involved in the expressions
for $\gamma^{(\nu,\bar\nu)}$ explicitly for some idealized situations,
which nevertheless should serve as starting point to consider more general
and/or realistic cases.
\section{Discussion}
\label{sec:discussion}
\label{sec:examples}

\subsection{Example of calculation of integrals}

\Eqs{Ix12final}{gammalinbladfinal} serve as the basis for the
calculation of the matrix $L$ using \Eq{Lxidentification}
in a number of useful cases. For illustrative purposes and a guide
to applications to realistic and/or
potentially important situations, here we evaluate explicitly
the integrals involved for some specific simple
cases of the background conditions.

We assume that $f_{x} \ll 1$ so that we can set
$(1 -f_{x}(E_{p^\prime})) \rightarrow 1$. Then
\beqa
\label{I12example}
I^{(x)}_1 & = & \frac{2}{\omega_\kappa}\left(\frac{1}{2\pi}\right)^5
\int\frac{d^3p}{2E_{p}} f_x(E_p) J_1(p,k)\,,\nonumber\\
I^{(x)}_2 & = & \frac{2}{\omega_\kappa}\left(\frac{1}{2\pi}\right)^5
\int\frac{d^3p}{2E_{p}} f_{x}(E_p) J_2(p,k)\,,
\eeqa
where
\beqa
\label{J12def}
J_1 & = & \int\frac{d^3p^\prime}{2E_{p^\prime}}
\frac{d^3\kappa^\prime}{2\omega_{\kappa^\prime}}
\delta^{(4)}(p + k - p^\prime - k^\prime)(p\cdot k^\prime)^2\,,\nonumber\\
J_2 & = & \int\frac{d^3p^\prime}{2E_{p^\prime}}
\frac{d^3\kappa^\prime}{2\omega_{\kappa^\prime}}
\delta^{(4)}(p + k - p^\prime - k^\prime)(p\cdot k)^2\,.
\eeqa
The evaluation of the integrals $J_{1,2}$ is straightforward, as shown
in \Appendix{sec:integralsJ}. Here we quote the results for
the particular cases of an ultrarelativistic or a non-relativistic
fermion background.
Although the idealizations and approximations we have made to arrive at these
formulas may limit their applicability to realistic situations,
the simplicity of these results can be used as a guide and benchmarks
when considering specific applications of practical interest.

\subsubsection{Ultrarelativistic background}

Specifically we assume that
\beq
\alpha_{f}, T,\omega_\kappa \gg m_f\,.
\eeq
As shown in \Appendix{sec:integralsJ}, in this case
\beq
\frac{1}{3}J_2 = J_1 = \frac{\pi}{6}\omega^2_\kappa p^2(1 - \cos\theta_p)^2\,,
\eeq
where $p = |\vec p|$. Then from \Eq{I12example},
\beq
\frac{1}{3}I^{(x)}_2 = I^{(x)}_1 =
\frac{\kappa}{36\pi^3}\int^\infty_0 dp p^3 f_x(p)\,,
\eeq
remembering that $\omega_\kappa = \kappa$.
For a completely degenerate $x$ background ($x = f$ or $\bar f$)
putting $f_x = \theta(p_{Fx} - p)$, where $p_{Fx}$ is the Fermi momentum,
\beq
\frac{1}{3}I^{(x)}_2 = I^{(x)}_1 =
\frac{\kappa}{36\pi^3}\frac{p^4_{Fx}}{4} \qquad \mbox{(Fermi gas)}\,.
\eeq
The Fermi momentum is given in terms of the number density $f_x$
of the background fermions by $p_{Fx} = (3\pi^2 n_x)^{\frac{4}{3}}$.
On the opposite side, for a classical background,
putting $f_x = e^{-\beta p}$, where $\beta$ is the
inverse temperature ($T$), gives
\beq
\frac{1}{3}I^{(x)}_2 = I^{(x)}_1 = \frac{\kappa T^4}{6\pi^3}
\qquad \mbox{(Classical gas)}\,.
\eeq
Using the above results in \Eq{gammalinbladfinal} we can consider some
specific example situations. For example, for a completely degenerate
$f$ gas (and no $\bar f$ particles),
\beq
\frac{1}{3}\gamma^{(\bar\nu)} = \gamma^{(\nu)} =
\frac{\kappa p^4_{Ff}}{144\pi^3 m^4_\phi} \qquad \mbox{(Fermi $f$ gas)}\,,
\eeq
or, for a completely degenerate $\bar f$ gas (and no $f$ particles),
\beq
\frac{1}{3}\gamma^{(\nu)} = \gamma^{(\bar\nu)} =
\frac{\kappa p^4_{F\bar f}}{144\pi^3 m^4_\phi} \qquad
\mbox{(Fermi $\bar f$ gas)}\,,
\eeq
while for a classical gas (equal number of $f$ and $\bar f$)
\beq
\gamma^{(\nu)} = \gamma^{(\bar\nu)} = \frac{2\kappa T^4}{3\pi^3 m^4_\phi}\,.
\eeq

\subsubsection{Nonrelativistic background}

Here we assume that
\beq
\omega_\kappa \gg m_f \gg T\,.
\eeq
As shown in \Appendix{sec:integralsJ}, in this case
\beq
\label{I12examplenr}
I^{(x)}_1 = \frac{1}{3}I^{(x)}_2 = \frac{\kappa m_f n_x}{48\pi}\,,
\eeq
where
\beq
n_x = 2\int\frac{d^3p}{(2\pi)^3} f_x(p)\,,
\eeq
is the total number density of $f$ or $\bar f$.
Thus, from \Eq{gammalinbladfinal},
\beqa
\gamma^{(\nu)} & = & \frac{1}{m^4_\phi}
\left(\frac{\kappa m_f}{48\pi}\right)(n_f + 3n_{\bar f})\,,\nonumber\\
\gamma^{(\bar\nu)} & = & \frac{1}{m^4_\phi}
\left(\frac{\kappa m_f}{48\pi}\right)(3n_f + n_{\bar f})\,.
\eeqa

\subsection{Generalizations}

As already mentioned in \Section{sec:introduction}, the method we have
followed, and the formulas we have obtained for the \emph{jump}
operators, can be applied with minor modifications to other model
interactions of potential interest. In particular, they can be applied
to study the effects of the non-forward scatering of neutrinos
when they propagate through a matter background
due to the standard weak interactions of the neutrinos with the
electrons and nucleons. For example, consider the contribution
from the electron background. In the local limit of the weak
interactions, the kinematics of the diagrams involved are similar
to those of the \emph{heavy $\phi$ limit} that we have assumed.
The corresponding jump operators would be given by formulas of the same
form as those in \Eqs{gammalinbladfinal}{Lxidentification}, with obvious
replacements. That is, $m_\phi \rightarrow m_W$, while the couplings $g_a$
would have the standard weak coupling $g$ as a common factor times another
factor that is the same for $\nu_{\mu,\tau}$ but different for $\nu_e$ due to
the charged-current interaction of the $\nu_e$ with the electrons. Similarly,
the integrals would involve the background electron number density, but
the specific kinematic factors involved must be determined by explicit
calculation. The details of the calculation would be similar to those
presented above.

In the phenomenological approaches to neutrino decoherence
in the references we have cited, the decoherence terms are considered
to be unknown phenomenological parameters, and the question of what
particular model (e.g., neutrino interactions and mixing)
can lead to one or another specific form of the decoherence matrices
is not considered. In contrast, for example, in the context of the model
we consider (one fermion $f$ and one scalar $\phi$), the matrices
$L^{(\nu,\bar\nu)}_{ab}$
would be diagonal of the form $diag(L^{(\nu,\bar\nu)}_{11},0,0)$
if only one neutrino couples to $f$ and $\phi$ so that only one coupling $g_a$
is non-zero (say $g_1$).
In models with more than one $f$ fermion, for example, it would be possible
to have more general diagonal $L$ matrices.
In any case, in the approach that we have initiated, the decoherence matrices
are determined in terms of the parameteres of the theory
(interactions, mixing, conditions of the evironment), and in principle
it allows to establish a link between the phenomenological form
of the decoherence matrices and the fundamental parameters
of the model.
\section{Conclusions and outlook}
\label{sec:conclusions}

In this work we have considered the damping effects in the
propagation of neutrinos in a background composed of a scalar particle
and a fermion with an interaction of the form given in \Eq{Lfnuphia},
due to the non-forward neutrino scattering processes.
Specifically, we calculated the contribution to the imaginary part of the
neutrino thermal self-energy arising from the non-forward neutrino
scattering processes in such backgrounds, from which the damping matrix
is determined. Since in this case the initial neutrino state is
depleted but does not actually disappear we have argued that
the damping matrix should be associated with decoherence effects.
Following this suggestion we have given a precise
prescription to determine the decoherence terms, as used in the context
of the master or Linblad equation, in terms of the damping terms
we have obtained from the calculation of the non-forward neutrino scattering
contribution to the imaginary part of the neutrino self-energy.
The main result is a well-defined formula for the ``jump'' operators in
that context, expressed in terms of integrals over the background
fermion distribution functions and the couplings constants of the interaction
of the neutrinos with the background particles in the model we consider.
The results can be useful in the context of Dark Matter-neutrino
interaction models in which the scalar and/or fermion constitute
the dark-matter, and can also serve to guide the application
to other models and/or situations that have been considered
recently using the Linblad equation (e.g., \cite{Coloma:2018idr}) in which the
decoherence effects in the propagation of neutrinos may be important.
For reference and guidance purposes we have evaluated 
the integrals involved explicitly for some conditions of the
background. Despite those simplifications the results illustrate
some features that can serve as a guide when considering more general cases
or situations not envisioned here.

The work of S. S. is partially supported by DGAPA-UNAM
(Mexico) Project No. IN103019.

\appendix
\section{Derivation of \Eq{nidentities}}
\label{sec:nidentities}
  
Here we show the details leading to \Eq{nidentities}. We will use the
fact that the $x's$ satisfy
\beq
x^\prime_\nu + x^\prime_f = x_\nu + x_f\,.
\eeq
We then have
\beqa
\label{Xaux}
X & \equiv &
\frac{1}{n_F(x_\nu)}e^{x_f} n_F(x_f) n_F(x^\prime_f) n_F(x^\prime_\nu)
\nonumber\\
& = & (e^{x_\nu} + 1)e^{x_f}n_F(x_f) n_F(x^\prime_f) n_F(x^\prime_\nu)
\nonumber\\
& = & (e^{x_\nu + x_f} + e^{x_f})n_F(x_f) n_F(x^\prime_f) n_F(x^\prime_\nu)
\nonumber\\
& = & (e^{x^\prime_\nu + x^\prime_f} +
e^{x_f})n_F(x_f) n_F(x^\prime_f) n_F(x^\prime_\nu)\,.
\eeqa
Using
\beq
e^x = \frac{1}{n_F(x)} - 1
\eeq
we now work the first factor,
\beqa
e^{x^\prime_\nu + x^\prime_f} + e^{x_f}
& = & \frac{1}{n_F({x^\prime}_\nu)n_F(x^\prime_f)} + \frac{1}{n_F(x_f)} -
\frac{1}{n_F(x^\prime_\nu)} - \frac{1}{n_F(x^\prime_f)}\,,
\eeqa
and using this in \Eq{Xaux} we get \Eq{nidentities}.
\section{Calculation of integrals $J_{1,2}$ in \Eq{J12def}}
\label{sec:integralsJ}

Since $J_{1,2}$ are a scalar integrals, we choose to do the integration in
the frame in which $p^\mu = (m_f,\vec 0)$ (the \emph{lab frame}). We
label the quantities in that frame with an asterisk,
$k^\mu = (\omega^\ast_k,\vec k^\ast)$ and similarly for $k^{\prime\mu}$,
and therefore
\beqa
J_1 & = & \int\frac{d^3k^{\ast\prime}}{2\omega^\ast_{k^\prime}}
\delta[(p + k - k^\prime)^2 - m^2_f]
\theta(m_f + \omega^\ast_k - \omega^\ast_{k^\prime})
(m_f\omega^\ast_{k^\prime})^2\nonumber\\
& = & \int\frac{d^3k^{\ast\prime}}{2\omega^\ast_{k^\prime}}
\delta[-2\omega^\ast_k \omega^\ast_{k^\prime}(1 - \cos\theta^\ast_{k^\prime}) +
2m_f(\omega^\ast_k - \omega^\ast_{k^\prime})]
\theta(m_f + \omega^\ast_k - \omega^\ast_{k^\prime})
(m_f\omega^\ast_{k^\prime})^2\,,
\eeqa
where $\theta^\ast_{k^\prime}$ is the angle between $\vec k^\ast$
and $\vec k^{\ast\prime}$. Carrying out with the integration over
$\cos\theta^\ast_{k^\prime}$ first, with the help of the $\delta$ function,
yields
\beq
\cos\theta^\ast_{k^\prime} = 1 -
\frac{m_f}{\omega^\ast_{k}\omega^\ast_{k^\prime}}
\left(\omega^\ast_{k} - \omega^\ast_{k^\prime}\right)\,,
\eeq
and
\beqa
\label{J1resultA}
J_1 & = &\frac{\pi m^2_f}{2\omega^\ast_k}
\int^{\omega^{\ast\prime}_{max}}_{\omega^{\ast\prime}_{min}}
d\omega^{\ast}_{k^\prime}\,
\omega^{\ast\,2}_{k^\prime}\nonumber\\
& = & \frac{\pi m^2_f}{6\omega^\ast_k}
\left(\omega^{\ast\prime\,3}_{max} - \omega^{\ast\prime\,3}_{min}\right)\,,
\eeqa
where the requirement that $-1 \le \cos\theta^\ast_{k^\prime} \le 1$ implies
\beqa
\label{omegaprimelabminmax}
\omega^{\ast\prime}_{min} & = &
\frac{m_f\omega^\ast_k}{m_f + 2\omega^\ast_k}\,,\nonumber\\
\omega^{\ast\prime}_{max} & = & \omega^\ast_k\,.
\eeqa
For $J_2$ we proceed similarly, with the replacement
$p\cdot k^\prime \rightarrow p\cdot k = m_f\omega^\ast_\kappa$
in the integrand, and thus,
\beqa
\label{J2resultA}
J_2& = & \frac{\pi m^2_f\omega^\ast_\kappa}{2}
\left(\omega^{\ast\prime}_{max} - \omega^{\ast\prime}_{min}\right)\,.
\eeqa

In order to use \Eqs{J1resultA}{J2resultA} in \Eq{I12example}, we express
$\omega^{\ast\prime}_{min}$ and $\omega^{\ast\prime}_{max}$ in terms
of $E_p$ and $|\vec p|$ by means of the relation
\beq
\label{omegalab}
\omega^\ast_k = \frac{1}{m_f}p\cdot k = \frac{\omega_\kappa E_p}{m_f}
(1 - v_p\cos\theta_p)\,,
\eeq
with $v_p = |\vec p|/E_p$. This allows the angular
integration in \Eq{I12example}
to be carried out in straightforward fashion, leaving only the integration
over $E_p$, which depends on the distribution function, to be performed.
As usual we can consider special cases for illustrative purposes.

\subsection{Ultrarelativistic background}

Specifically we assume that
\beq
\alpha_f, T,\omega_\kappa \gg m_f\,.
\eeq
In this case,
\beq
\label{wprimeastminzero}
\omega^{\ast\prime}_{min} = 0\,,
\eeq
and therefore
\beq
J_1 = \frac{\pi}{6}(m_f\omega^\ast_k)^2 \rightarrow
\frac{\pi}{6}\omega^2_\kappa |\vec p|^2(1 - \cos\theta_p)^2\,.
\eeq
Then, denoting $p = |\vec p|$, from \Eq{I12example},
\beqa
I^{(x)}_1 & = &
\frac{2}{\omega_\kappa}\left(\frac{1}{2\pi}\right)^5
\frac{\pi^2\omega^2_\kappa}{6}
\frac{8}{3}\int^\infty_0 dp p^3 f_x(p)\nonumber\\
& = & \frac{\kappa}{36\pi^3}\int^\infty_0 dp p^3 f_x(p)\,,
\eeqa
remembering that $\omega_\kappa = \kappa$. Similarly,
\beq
J_2 = \frac{\pi}{2}(m_f\omega^\ast_k)^2 = 3J_1
\eeq
and therefore
\beq
I^{(x)}_2 = \frac{\kappa}{12\pi^3}\int^\infty_0 dp p^3 f_{x}(p)\,,
\eeq

\subsubsection{Completely degenerate background}

For a completely degenerate $x$ background ($x = f$ or $\bar f$)
putting $f_x = \theta(p_{Fx} - p)$, where $p_{Fx}$ is the Fermi momentum,
\beqa
I^{(x)}_1 & = & \frac{\kappa}{36\pi^3}\frac{p^4_{Fx}}{4}\,,\nonumber\\
I^{(x)}_2 & = & \frac{\kappa}{12\pi^3}\frac{p^4_{Fx}}{4}\,.
\eeqa
The Fermi momentum is given in terms of the number density $f_x$
of the background fermions by $p_{Fx} = (3\pi^2 n_x)^{\frac{4}{3}}$.

\subsubsection{Classical background}

Putting $f_x = e^{-\beta p}$, where $\beta$ is the
inverse temperature ($T$), gives
\beqa
I^{(x)}_1 & = & \frac{6\kappa}{36\pi^3 \beta^4} = \frac{\kappa T^4}{6\pi^3}\,,
\nonumber\\
I^{(x)}_2 & = & 3I^{(x)}_1 = \frac{\kappa T^4}{2\pi^3}\,.
\eeqa

\subsection{Nonrelativistic background}

Here we assume that
\beq
\omega_\kappa \gg m_f \gg T\,.
\eeq
In this case, from \Eq{omegalab},
\beq
\omega^\ast_k = \omega_\kappa\,,
\eeq
and we have \Eq{wprimeastminzero} once again. Thus,
\beq
J_1 = \frac{\pi}{6}(m_f\omega^\ast_\kappa)^2 \rightarrow
\frac{\pi}{6}m^2_f\omega^2_\kappa\,,
\eeq
and similarly we get
\beq
J_2 = 3J_1\,,
\eeq
in this case as well. Thus from \Eq{I12example}, we arrive
at the result quoted in \Eq{I12examplenr}.


\begin{thebibliography}{99}
\bibitem{Mangano:2006mp}
    G.~Mangano, A.~Melchiorri, P.~Serra, A.~Cooray
	    and M.~Kamionkowski,
    \emph{Cosmological bounds on dark matter-neutrino interactions},
    Phys. Rev. D 74, 043517 (2006)
    [arXiv:astro-ph/0606190].

\bibitem{Binder:2016pnr}
    T.~Binder, L.~Covi, A.~Kamada, H.~Murayama,
	    T.~Takahashi and N.~Yoshida,
    \emph{Matter Power Spectrum in Hidden Neutrino Interacting
	    Dark Matter Models: A Closer Look at the Collision Term},
    JCAP 1611, 043 (2016)
    [arXiv:1602.07624].

\bibitem{Primulando:2017kxf}
    R.~Primulando and P.~Uttayarat,
    \emph{Dark Matter-Neutrino Interaction in Light of Collider
	    and Neutrino Telescope Data},
    JHEP 1806, 026 (2018)
    [arXiv:1710.08567].
    
\bibitem{Campo:2017nwh}
    A.~Olivares-Del Campo, C.~Bœhm, S.~Palomares-Ruiz
	    and S.~Pascoli,
    \emph{Dark matter-neutrino interactions through the lens
	    of their cosmological implications},
    Phys. Rev. D 97, 075039 (2018)
    [arXiv:1711.05283].

\bibitem{Brune:2018sab} 
  T.~Brune and H.~P\"as,
\emph{Massive Majorons and constraints on the Majoron-neutrino coupling},
  Phys.\ Rev.\ D {\bf 99}, 096005 (2019)
  [arXiv:1808.08158 [hep-ph]].

\bibitem{Franarin:2018gfk}
    T.~Franarin, M.~Fairbairn and J.~H.~Davis,
    \emph{JUNO Sensitivity to Resonant Absorption of Galactic Supernova
	    Neutrinos by Dark Matter},
    [arXiv:1806.05015].

\bibitem{Dev:2019anc} 
  P.~S.~Bhupal Dev {\it et al.},
 \emph{ Neutrino Non-Standard Interactions: A Status Report},
  arXiv:1907.00991 [hep-ph].

\bibitem{Pandey:2018wvh} 
  S.~Pandey, S.~Karmakar and S.~Rakshit,
  \emph{Interactions of Astrophysical Neutrinos with Dark Matter:
  A model building perspective},
  JHEP {\bf 1901}, 095 (2019)
  [arXiv:1810.04203 [hep-ph]].

\bibitem{Karmakar:2018fno} 
  S.~Karmakar, S.~Pandey and S.~Rakshit,
 \emph{Are We Looking at Neutrino Absorption Spectra at IceCube?},
  arXiv:1810.04192 [hep-ph].
  
\bibitem{Nieves:2018vxl}
    J.\ F.\ Nieves and S.\ Sahu,
    \emph{Neutrino effective potential in a fermion and scalar background},
    Phys. Rev. D 98, 063003 (2018)
    [arXiv:1808.01629].

\bibitem{nsnuphidamp}
    J.\ F.\ Nieves and S.\ Sahu,
    \emph{Neutrino damping in a fermion and scalar background},
    Phys. Rev. D 99, 095013 (2019)
    [arXiv:1812.05672]

\bibitem{Coloma:2018idr} 
  P.~Coloma, J.~Lopez-Pavon, I.~Martinez-Soler and H.~Nunokawa,
  \emph{Decoherence in Neutrino Propagation Through Matter,
                    and Bounds from IceCube/DeepCore},
  Eur.\ Phys.\ J.\ C {\bf 78}, no. 8, 614 (2018)
 [arXiv:1803.04438 [hep-ph]].

\bibitem{Carpio:2017nui} 
  J.~A.~Carpio, E.~Massoni and A.~M.~Gago,
 \emph{Revisiting quantum decoherence for neutrino oscillations in matter
                  with constant density},
  Phys.\ Rev.\ D {\bf 97}, no. 11, 115017 (2018)
 [arXiv:1711.03680 [hep-ph]].

\bibitem{Oliveira:2014jsa} 
  M.~M.~Guzzo, P.~C.~de Holanda and R.~L.~N.~Oliveira,
 \emph{Quantum Dissipation in a Neutrino System Propagating
               in Vacuum and in Matter},
  Nucl.\ Phys.\ B {\bf 908}, 408 (2016)
[arXiv:1408.0823 [hep-ph]].

\bibitem{Fogli:2007tx} 
  G.~L.~Fogli, E.~Lisi, A.~Marrone, D.~Montanino and A.~Palazzo,
 \emph{Probing non-standard decoherence effects with
               solar and KamLAND neutrinos},
  Phys.\ Rev.\ D {\bf 76}, 033006 (2007)
[arXiv:0704.2568 [hep-ph]].

\bibitem{Capolupo:2018hrp} 
  A.~Capolupo, S.~M.~Giampaolo and G.~Lambiase,
\emph{Decoherence in neutrino oscillations, neutrino nature and CPT violation},
  Phys.\ Lett.\ B {\bf 792}, 298 (2019)
  [arXiv:1807.07823 [hep-ph]].

\bibitem{Chu:2018gxk} 
  X.~Chu, B.~Dasgupta, M.~Dentler, J.~Kopp and N.~Saviano,
 \emph{Sterile neutrinos with secret interactions—cosmological discord?},
  JCAP {\bf 1811}, 049 (2018)
  [arXiv:1806.10629 [hep-ph]].

\bibitem{Daley:2014fha} 
  A.~J.~Daley,
\emph{Quantum trajectories and open many-body quantum systems},
  Adv.\ Phys.\  {\bf 63}, no. 2, 77 (2014)
  [arXiv:1405.6694 [quant-ph]].

\bibitem{Weinberg:2011jg} 
  S.~Weinberg,
 \emph{Collapse of the State Vector},
  Phys.\ Rev.\ A {\bf 85}, 062116 (2012)
  [arXiv:1109.6462 [quant-ph]].

\bibitem{pearle} P. Pearle,
\emph{Simple derivation of the Lindblad equation},
Eur. J. Phys. 33, 805 (2012),
  [arXiv:1204.2016 [math-ph]].

\bibitem{Plenio:1997ep} 
  M.~B.~Plenio and P.~L.~Knight,
\emph{The Quantum jump approach to dissipative dynamics in quantum optics},
  Rev.\ Mod.\ Phys.\  {\bf 70}, 101 (1998)
  [quant-ph/9702007].

\bibitem{Lieu:2019cev} 
  S.~Lieu,
\emph{Non-Hermitian Majorana modes protect degenerate steady states},
  Phys.\ Rev.\ B {\bf 100}, 085110 (2019)
  [arXiv:1904.07481 [cond-mat.mes-hall]].

\bibitem{footnote1}
Strictly speaking this is correct in the massless neutrino limit,
which in practice is a valid approximation in the limit that the neutrino
mass can be neglected in the calculation of the relevant diagrams.

\bibitem{footnote2}
Actually, under the isotropy assumption
$V^{(u)},V^{(t)}$ depend on $\omega$ and $\kappa$ but not on
the direction of $\vec\kappa$.

\bibitem{Kadanoff:1962}
L. P. Kadanoff and G. Byam,
\emph{Quantum Statistical Mechanics}, p. 37 (Benjamin, New York, 1962).

\bibitem{footnote3}
The author of Ref.\ \cite{Lieu:2019cev} states it clearly like this:
\emph{The Lindblad master equation (5) lends itself to a
convenient  physical  interpretation  known  as  the  quantum
stochastic wavefunction approach [46, 47]:  In a time step $dt$,
a system prepared in a pure state will either evolve
coherently according to the non-Hermitian effective Hamiltonian
$H_{eff}$ or a ``quantum jump event'' will decohere the
system by moving a pure state from $|\psi\rangle$ to
$L_i|\psi\rangle$. Averaging over all such trajectories will produce the same
expectation values as formally solving the Lindblad master
equation for the evolution of the density matrix.}
\end{thebibliography}

\end{document}